\documentclass[12pt,twoside]{article}
\usepackage{a4}\usepackage{amsfonts}\usepackage{amsfonts}
\begin{document}
\begin{titlepage}
\begin{center}
\textbf{\Large{ Boundary dynamics and multiple reflection
expansion for Robin boundary conditions}}
\vspace{1cm}

\textbf{M.~Bordag$^\star$\footnotemark[1],
H.~Falomir$^\dag$\footnotemark[2],
E.M.~Santangelo$^\dag$\footnotemark[3],
D.V.~Vassilevich$^\star$\footnotemark[4]}

\vspace{7ex}

{$^\star$University of Leipzig, Institute for Theoretical Physics\\
  Augustusplatz 10/11, 04109 Leipzig, Germany}

\vspace{2ex}
{$^\dag$Departamento de F\'{\i}sica, Universidad Nacional de La Plata\\
  C.C.67, 1900 La Plata,Argentina}
\footnotetext[1]{e-mail: Michael.Bordag@itp.uni-leipzig.de}
\footnotetext[2]{e-mail: falomir@obelix.fisica.unlp.edu.ar}
\footnotetext[3]{e-mail: mariel@obelix.fisica.unlp.edu.ar}
\footnotetext[4]{e-mail: Dmitri.Vassilevich@itp.uni-leipzig.de,
On leave from V.A.~Fock Institute of Physics, St.Petersburg University,
198904 St.Petersburg, Russia}
\end{center}

\vfill

\begin{abstract}
In the presence of a boundary interaction,  Neumann boundary
conditions should be modified to contain a function $S$ of the
boundary fields: $(\nabla_N +S)\phi =0$. Information on quantum
boundary dynamics is then encoded in the $S$-dependent part of the
effective action. In the present paper we extend the multiple
reflection expansion method to the Robin boundary conditions
mentioned above, and calculate the heat kernel and the effective
action (i) for constant $S$, (ii) to the order $S^2$ with an
arbitrary number of tangential derivatives. Some applications to
symmetry breaking effects, tachyon condensation and brane world are
briefly discussed.
\end{abstract}
\end{titlepage}
\section{Introduction}
Recent years have seen a considerable increase of the interest in
the relations between bulk and boundary dynamics. One of the most
exciting applications of the subject is the tachyon condensation in
open string theory. Modern interest in this mechanism of symmetry
breaking is connected with the papers by Sen and Zwiebach
\cite{sen}, though certain ideas were developed much earlier (see,
e.g., \cite{tc-old}). Now, many different methods are being used in
this field. An extensive literature survey can be found in
\cite{rev-tc}. The sigma model approach \cite{Tseytlin:2001mt} is
probably the closest one to the technique of the present paper. The
tachyon field enters the boundary term of the open string action,
and, therefore, modifies the open string boundary conditions.

In the present paper we deal with a scalar theory on a manifold
$\cal M$, ${\rm dim}\, {\cal M}=m$. Let the classical action be of
the form:
\begin{equation}
{\cal S}(\phi )=\int_{\cal M} d^mx\left( (\nabla \phi )^2 +V(\phi
)\right) +\int_{\partial {\cal M}} d^{m-1}x\tilde V(\phi )\,,
\label{clac}
\end{equation}
with two so far arbitrary potentials $V$ and $\tilde V$. If $\cal M$
is an open string world surface, and if $\phi $ is the string
coordinate\footnote{The coordinates $X^\mu$ are scalars from the
world surface point of view.} $X^\mu$, then the boundary potential
$\tilde V$ may be identified with the boundary tachyon: $\tilde
V(\phi ):=T(X)$.

Let us split $\phi$ into its background part $\bar\phi$ and the
quantum fluctuations $\varphi$: $\phi =\bar\phi +\varphi$. To
calculate the one-loop effective action we have to keep the part of
${\cal S}$ which is quadratic in $\varphi$:
\begin{equation}
{\cal S}_2=\int_{\cal M} d^mx\,\varphi D \varphi +\int_{\partial
{\cal M}} d^{m-1}x \,\varphi (-\nabla_N -S)\varphi
\,,\label{quaction}
\end{equation}
where $\nabla_N$ is covariant derivative with respect to the inward
pointing unit normal. $D$ is a second order partial differential
operator which depends on $\bar\phi$, $S=-\frac 12 \tilde
V''(\bar\phi)$; prime denotes differentiation with respect to $\phi$.
Boundary conditions for the fluctuations $\varphi$ follow from the
requirement that the boundary part of the action (\ref{quaction})
vanishes. These can be either Dirichlet $\varphi\vert_{\partial
{\cal M}}=0$ or Robin
\begin{equation}
(\nabla_N +S)\varphi \vert_{\partial {\cal M}}=0 \label{rbc}
\end{equation}
boundary conditions. Both sets of boundary conditions ensure also
the Hermiticity of the operator $D$. In the present paper we will
consider Robin boundary conditions only. We will be interested in
the dependence on the function $S$ of the heat kernel coefficients,
the trace of the heat kernel, and the one-loop effective action.

The brane-world scenario \cite{bw} usually assumes that there is
an interaction of the bulk fields which is confined on a surface
$\Sigma$. Many essential features of such interactions can be
described by the action (\ref{clac}) where in the second term
one integrates over the surface $\Sigma$. One-loop quantum corrections are
then given by the determinant of the operator $D$ subject to the following
matching conditions on $\Sigma$,
\begin{eqnarray}
&&\varphi_+=\varphi_- \,,\nonumber \\
&&-(\nabla_N\varphi)_++(\nabla_N\varphi)_-+\frac 12
\tilde V(\bar\phi)\varphi=0\,,
\label{mcon}
\end{eqnarray}
where the subscripts ``+'' and ``-'' denote limiting values on $\Sigma$
from the two sides of the surface. It has been demonstrated in
\cite{Bordag:1999ed,Gilkey:2001mj} that if the background fields
are symmetric under the reflection about $\Sigma$, the $\tilde V$-dependent
part of the heat kernel (and, consequently, that of the effective action)
is indeed described by the Robin boundary value problem (\ref{rbc})
with $S=-\frac 14 \tilde V(\bar \phi)$. Hence, the results for the
effective action which we will obtain in this paper are valid also
for the brane-world scalar field (though they will be, of course,
modified by the presence of the background curvature).

These two examples -- strings and brane-world -- are the main
physical motivations for our study. For this reason, our explicit
calculations will be carried out with an emphasis on dimensions two
and five. However, since the main purpose of this paper is to
develop a technique, we will not go too far in the applications (and
will not give a more detailed literature survey). We reserve these
subjects for future publications.

The heat kernel technique is by now a standard tool of quantum field
theory. It allows to extract divergences and anomalies in a very
efficient way, and to represent the one loop effective action in a
nice geometric form. More details on related mathematics and physical
applications may be found in the books \cite{books}. Moreover, a
knowledge of the effective potential allows one to study symmetry
breaking effects with the formation of a (boundary) condensate. The
imaginary part of the quantum potential tells us about vacuum
instability, while the derivative part of the effective action is
related to the momentum dependence of certain diagrams.

In previous papers \cite{Bordag:1999ed,mrnos}, we showed that the multiple
reflection expansion \cite{mr} is a powerful tool for studying the
asymptotics of the heat kernel, and applied it to some classical
boundary problems, as well as singular potentials for second order
operators on compact manifolds with spherical symmetry.

In the present paper, we apply the same technique to second order
differential operators on flat $m$-dimensional manifolds, acting on
functions that satisfy, at the $(m-1)$-dimensional boundary, the
conditions in equation (\ref{rbc}).

In Section \ref{21}, the general form of the smeared boundary heat
kernel coefficients is obtained. In the particular case of a constant
field $S$, and after taking the smearing function to unity, the
resulting series is shown to be summable, and the complete boundary
contribution to the trace of the heat kernel is then obtained for
any dimension $m$ in Section \ref{22}.

Section \ref{23} contains the multiple reflection expansion study of the
boundary contribution to the trace of the heat kernel in the case of
a field $S$ depending on the boundary coordinates. In this
situation, each order in the multiple reflection expansion is shown
to give, not only the asymptotic contributions, but also the
non-asymptotic ones.

Using these results, the boundary one-loop effective Lagrangian is
evaluated for constant $S$ in Section \ref{31}. The boundary one-loop
effective kinetic energy for a general $S(z)$ is studied in Section
\ref{32}.

In Appendix \ref{A1} we show that our results for the trace of the
heat kernel can be extended to curved boundaries. Appendix \ref{A2}
contains some details on the treatment of possible negative
eigenvalues. Explicit expressions for the boundary part of the
Coleman--Weinberg potential for massive fields are given in Appendix
\ref{A3}.

\section{Heat kernel}
\subsection{Multiple reflection expansion}
\label{21} In this paper we are interested in the dependence of the
heat kernel and the effective action on boundary values of the
background field $\bar \phi$, through the function $S$ in the
boundary conditions (\ref{rbc}). We perform actual calculations for
a simple flat geometry. Our results must then be understood as an
approximation to more generic situations.

Let $\cal M$ be a flat, $m$-dimensional half space, and $V(\phi )=0$
in equation (\ref{clac}) (then $D$ is free scalar Laplacian).
The heat kernel $K(x,y;t)$ is defined as a solution of the heat
equation
\begin{equation}
(\partial_t + D_x )K(x,y;t)=0 \label{heq}
\end{equation}
with the initial condition $K(x,y;0)=\delta(x-y)$ inside the
manifold. It must be supplemented with (Robin) boundary conditions
when its first argument belongs to the boundary.

The heat kernel for Robin boundary conditions satisfies the Dyson equation
\begin{equation}
K(x,y;t)=K_N(x,y;t)+\int_0^t ds\int_{\partial M} dz
K_N(x,z;s) S(z) K(z,y;t-s) \,, \label{Deq}
\end{equation}
where $K_N$ is the heat kernel for Neumann boundary conditions
($S=0$). It admits a solution in terms of a power series in $S$
\begin{eqnarray}
&&K(x,y;t)=K_N(x,y;t) + \sum_{n=1}^\infty \int\limits_0^t ds_n
\int\limits_0^{s_n}ds_{n-1} \dots \int\limits_0^{s_2}ds_1
\int\limits_{\partial M} dz_n
\dots \int\limits_{\partial M} dz_1 \nonumber
\\
&& \quad \times K_N(x,z_n;t-s_n)S(z_n)K_N(z_n,z_{n-1};s_n-s_{n-1})
\dots S(z_1)K_N(z_1,y;s_1) \,,\label{solint}
\end{eqnarray}
which is nothing but a multiple reflection representation for the
heat kernel. It is, of course, equivalent to the multiple reflection
representation for the propagator \cite{mr}.

A very simple (and also very naive) way to derive the equations
(\ref{Deq}) and (\ref{solint}) is to consider the boundary
interaction term $\int_{\partial M}\varphi S \varphi$ as a
perturbation to the Neumann problem, represent the heat kernel as
$K(x,y;t)=(x| \exp (-t(D+S\delta_{\partial M}) |y)$ and expand the
exponential in a power series of $S$. This derivation ignores
completely all existence and convergence requirements, but reproduces
correctly the combinatorics of the expansion. Similar arguments fail
for other types of perturbations.
 The equation (\ref{solint}) can be also derived from the
corresponding equation in \cite{Bordag:1999ed} using equivalence to
semi-transparent boundaries. Note that, after suitable modifications
in the multiple reflection expansion, $S$ may be replaced by a
general differential operator \cite{Kummer:2000ae}.

The representation (\ref{solint}) was used in the context of singular
potentials in \cite{Bordag:1999ed} to prove some general properties
of the heat kernel and also in \cite{Moss:2000gv} to calculate the
heat trace asymptotics.

Consider the smeared heat kernel
\begin{equation}
K(f,t)=\int_{\cal M} d^m x f(x) K(x,x;t) \,.\label{smhk}
\end{equation}
This is a mathematically consistent way to deal with the distributional
nature of the heat kernel diagonal. Considering just the integrated
heat kernel ($f=1$) is insufficient for many physical applications
where one needs local quantities (as, e.g., the trace anomaly).
Let $x=(z,r)$, where $z$ is a coordinate on the boundary, and
$r$ is a normal coordinate.
Let us expand the smearing function $f$ in a Taylor series in $r$:
\begin{equation}
f(z,r)=\sum_{p=0}^{\infty} \frac {r^p}{p!} f^{(p)}(z)
\,,\label{serinr}
\end{equation}
where $f^{(p)}(z)$ denotes the boundary value of the $p$-th normal
derivative. By substituting (\ref{solint}), (\ref{serinr}) in
(\ref{smhk}) and integrating over $r$ we obtain
\begin{eqnarray}
&&K(f,t)=\sum_{n=0}^\infty \sum_{p=0}^{\infty} \int_{\partial M} dz\
f^{(p)}(z)  \frac {2^p}{p!} \Gamma\left( \frac{1+p}2 \right)
\int_0^t ds_n \int_0^{s_n}ds_{n-1} \dots \int_0^{s_2}ds_1
\nonumber \\
&& \qquad \times \left( \frac{(t-s_n)s_1}{t-s_n+s_1}
\right)^{\frac{1+p}2} \int_{\partial M} dz_n \dots \int_{\partial M}
dz_1 K_N(z,z_n;t-s_n)S(z_n)
\nonumber \\
&& \qquad \times K_N(z_n,z_{n-1};s_n-s_{n-1}) \dots
S(z_1)K_N(z_1,z;s_1) \,,\label{main}
\end{eqnarray}
where we have used that
\begin{equation}
K_N(x,z_i;s)=\exp (-r^2/4s) K_N(z,z_i;s),\qquad x=(z,r)\,.
\label{norm}
\end{equation}
To study the Casimir interaction between two Robin boundaries
\cite{Romeo:2000wt}  one must replace (\ref{norm}) by the Neumann heat kernel
in the strip (which can be also presented in a closed form).

To evaluate (\ref{main}) in the case of a general, $z$-dependent $S$,
it will prove convenient to make a Fourier transformation
\begin{eqnarray}
&&K_N(z_i,z_{i-1};s)=\int \frac{d^{m-1}k_i}{(2\pi )^{(m-1)/2}}
\frac 1{\sqrt{\pi s}} \exp (-sk_i^2 -i(z_i-z_{i-1})k ) \nonumber \\
&&S(z_j)=\int d^{m-1}\tilde k_j \tilde S (\tilde k_j)\exp (i\tilde
k_jz_j)
\nonumber \\
&&\tilde S (\tilde k_j)=\int \frac{d^{m-1}z_j}{(2\pi )^{m-1}}
S(z_j)\exp (-i\tilde k_jz_j) \,.\label{Four}
\end{eqnarray}
However, we will first treat the case of a constant field, where it
is not necessary to go to the Fourier-conjugate space.

\subsection{Heat kernel for constant $S$}
\label{22}

In this case, an individual term in the double sum in (\ref{main})
reads:
\begin{eqnarray}
&&\int_{\partial {\cal M}} dz\ f^{(p)}(z)S^n  \frac {2^p}{p!}
\Gamma\left( \frac{1+p}2 \right) \int_{\partial {\cal M}} \dots
\int_{\partial {\cal M}} dz_1\dots dz_n \pi^{-\frac{n+1}2} \nonumber \\
&&\ \ \times \int_0^t ds_n \int_0^{s_n}ds_{n-1} \dots
\int_0^{s_2}ds_1 \left( \frac{(t-s_n)s_1}{t-s_n+s_1}
\right)^{\frac{1+p}2} \int \prod_{i=0}^n \frac{d^{m-1}k_i}{(2\pi
)^{m-1}}
\nonumber \\
&&\ \ \times [(t-s_n)(s_n-s_{n-1})\dots s_1]^{-\frac 12}
\exp (-(t-s_n)k_n^2 -\dots -s_1k_0^2) \nonumber \\
&&\ \ \times \exp (-ik_n(z-z_n)-ik_{n-1}(z_n-z_{n-1}-\dots
-ik_0(z_1-z))\,. \label{nodir1}
\end{eqnarray}
The integration over $z_i$ can be easily performed; it gives as a
result
\begin{equation}
\left[ (2\pi )^{m-1} \right]^n \delta (k_n-k_{n-1}) \delta
(k_{n-1}-k_{n-2}) \dots \delta (k_1-k_0)\,.
\end{equation}
These $\delta$-functions can be used to integrate over $k_i$, $i\ne
0$. Integration over $k_0$ gives then $(\pi /t)^{(m-1)/2}$. After
the change of variables $\alpha_1 =s_1/t $,
$\alpha_2=(s_2-s_1)/t$,\dots , $\alpha_0=1-s_n/t$, the integration
over $\{ \alpha \}$ reduces to $I\left( \frac p2 ,-\frac
{1+p}2,-\frac 12 ,\dots , -\frac 12 \right)$, where $I$ is a
particular value of the following integral
\begin{eqnarray}
&&I(A_0,A_1,\dots ,A_n)=\int_{\sum \alpha_i=1} \prod d\alpha_i
(\alpha_0\alpha_1)^{A_0}(\alpha_0+\alpha_1)^{A_1}
\prod_{i=2}^n \alpha_i^{A_i} \nonumber \\
&&\qquad =\frac{ \Gamma (2A_0+A_1+2) \Gamma(A_0+1)^2 \prod_{i=2}^n
\Gamma (A_i+1)}{\Gamma (2A_0+2) \Gamma (A_0 +\sum_{i=0}^n A_i +n
+1)}\,. \label{Int}
\end{eqnarray}
The final result for (\ref{nodir1}) reads:
\begin{equation}
\frac {1}{(4\pi )^{\frac{m-1}2}}\int_{\partial {\cal M}} dz\
f^{(p)}(z) S^n t^{\frac 12 (n+p-m+1)} \frac{2^{-p-1}}{\Gamma \left(
\frac{n+p}2 +1 \right)}\,. \label{nodir2}
\end{equation}

Note that each order in reflections gives one complete order in
powers of $t$.

For $n+p\le 4$, this result can be checked against the expressions of
\cite{BG90,Branson:1997cm,Branson:1999jz}. For $p=0$ we reproduce
the heat kernel expansion for a delta-potential in one dimension
\cite{blinder}.

The power series with individual terms given by
(\ref{nodir2}) can be summed up to give a closed
expression for the heat kernel. Taking the smearing function, $f=1$,
we obtain for the trace of the heat kernel
\begin{equation}
K(t)=\frac {V}{2(4\pi t)^{\frac{m-1}2}} \left( e^{S^2t} {\rm erf}
(S\sqrt t ) +e^{S^2t} \right)\,, \label{hkerf}
\end{equation}
where $V$ is the (infinite) volume of the boundary, and ${\rm erf}
(S\sqrt t)$ is the error function
\[
{\rm erf} (S\sqrt t )=\frac{2}{\sqrt \pi}\int_0^{S\sqrt t}d\xi
\,e^{-\xi^2}\,.
\]

 In particular, for $m=2$, equation (\ref{hkerf}) can be seen to coincide
with the trace of the heat kernel given, for example, in reference
\cite{carslaw}.

\subsection{Expansion to the order of two reflections for a $z$-dependent $S$}
\label{23}

As before, we will take $f=1$. The term corresponding to no
reflection ($n=0$) is independent of $S$. It is then given by
equation (\ref{nodir2}), with $p=0$, $f=1$ and $n=0$.

As for the one-reflection contribution, from equation (\ref{nodir1})
it can easily be seen to be

\begin{equation}
K_1=\frac{2^{1-m}}{\pi^\frac{m}{2}} t^{\frac
{(2-m)}{2}}\int_{\partial {\cal M}} dz S(z)=
\frac{2^{1-m}}{\pi^\frac{m}{2}} t^{\frac {(2-m)}{2}} \tilde S(\tilde
k=0)\,. \label{hk1}
\end{equation}

Now, consider the term with $n=2$, $p=0$ and $f=1$ in (\ref{main}):
\begin{eqnarray}
&&\int_{\partial {\cal M}} dz\ f(z)\int_{\partial {\cal M}}
\int_{\partial {\cal M}} dz_1\, dz_2 \frac 1\pi \int_0^t
ds_2\int_0^{s_2}ds_1 \left( \frac{(t-s_2)s_1}{t-s_2+s_1}
\right)^{\frac{1}2}
\nonumber \\
&&\ \ \times [(t-s_2)(s_2-s_1)s_1]^{-\frac 12} \int
\frac{dk_1dk_2dk_3d\tilde k_1d\tilde k_2}{(2\pi)^{3(m-1)}} \tilde
S(\tilde k_1) \tilde S(\tilde k_2)
\nonumber \\
&&\ \ \times
\exp (-k_1^2(t-s_2)-k_2^2(s_2-s_1)-k_3^2s_1) \nonumber \\
&&\ \ \times \exp( -ik_1(z-z_2)-ik_2(z_2-z_1)-ik_3(z_1-z)+i\tilde
k_1z_1 +i\tilde k_2z_2)\,. \label{S21}
\end{eqnarray}
Integration over $z_1$ and $z_2$ gives rise to $(2\pi )^{2(m-1)}
\delta (k_2-k_3+\tilde k_1) \delta (k_1-k_2+\tilde k_2)$. These
$\delta$-functions are then used to integrate over $k_1$ and $k_3$.
Next, we shift $k_2$ to complete the square in the exponential and
integrate over (shifted) $k_2$. We thus obtain:

\begin{eqnarray}
&&\int_{\partial {\cal M}} dz\  \frac 1\pi \int_0^t
ds_2\int_0^{s_2}ds_1 [(t-s_2+s_1)(s_2-s_1)]^{-\frac 12}
\nonumber \\
&&\ \ \times \frac 1{(4\pi t)^{\frac{m-1}2}} \int d\tilde
k_1d\tilde k_2 \tilde S(\tilde k_1) \tilde S(\tilde k_2) \exp
(iz(\tilde k_1 +
\tilde k_2) \nonumber \\
&&\ \ \times \exp \left( -\frac 1t (\tilde k_1^2 (t-s_1)s_1 +\tilde
k_2^2(t-s_2)s_2 +2\tilde k_1\tilde k_2 (t-s_2)s_1 )\right)\,.
\end{eqnarray}

The integration over $dz$ can now be performed. This gives

\[ \frac 1\pi \int_0^t ds_2\int_0^{s_2}ds_1
[(t-s_2+s_1)(s_2-s_1)]^{-\frac 12}\ \times \frac 1{(4\pi
t)^{\frac{m-1}2}} \]\[\int d^{m-1}k\  \tilde S(k) \tilde S(-k)
\exp \left( -\frac{k^2}{t}((s_2-s_1)(t-s_2+s_1)) \right) \]

or, after exponentiating the first factor

\[ \frac {1}{\pi(4\pi
t)^{\frac{m-1}2}} \int d^{m-1}k\ \tilde S(k) \tilde S(-k)\int_0^t
ds_2\int_0^{s_2}ds_1\ \times\]\[ \int_0^{\infty}\ dz\
\frac{z^{-\frac12}}{\Gamma(\frac12)}\exp \left( -(z+\frac{k^2}{t})\
(s_2-s_1)(t-s_2+s_1) \right) \,.\]

Calling $s_1^{'}=s_2-s_1$, and integrating by parts in $s_2$, one
gets

\[ \frac {t^2}{\pi(4\pi
t)^{\frac{m-1}2}} \int d^{m-1}k\ \tilde S(k) \tilde S(-k)
\int_0^{\infty}\ dz\
\frac{z^{-\frac12}}{\Gamma(\frac12)}\]\[\int_0^1\ ds\exp \left(
-(z+\frac{k^2}{t})t^2\ s(1-s) \right)\,. \]

After performing the integral in $z$, this last expression gives

\begin{equation} \frac {t^{\frac{3-m}{2}}}{\pi(4\pi)^{\frac{m-1}2}} \int
d^{m-1}k\ \tilde S(k) \tilde S(-k) \int_0^1\ ds\
\frac{(1-s)^{\frac12}}{s^{\frac12}}\  \exp \left( -tk^2 s(1-s)
\right)\,. \label{hk2}
\end{equation}

Using its transformation properties under $s\rightarrow 1-s$, the
last integral can be rewritten as
\[
\int_0^1\ ds\ \frac{(1-s)^{\frac12}}{s^{\frac12}}\  \exp \left( -tk^2
s(1-s) \right)=\frac12\,\int_0^1\ ds\
\frac{1}{(1-s)^{\frac12}s^{\frac12}}\  \exp \left( -tk^2 s(1-s)
\right)\]
\[
=\int_0^{\frac14}\ ds\ \frac{1}{(1-4s)^{\frac12}s^{\frac12}}\  \exp
\left( -tk^2 s \right)=\frac{\pi}{2}\exp \left( -t
\frac{k^2}{8}\right)I_{0}(t \frac{k^2}{8})\,,\] where $I_{0}(x)$ is
the modified Bessel function of order $0$.

So, after replacing this into equation (\ref{hk2}), the contribution
of two reflections to the trace of the heat kernel is seen to be

\begin{equation} K_2=\frac {t^{\frac{3-m}{2}}}{2(4\pi)^{\frac{m-1}2}} \int
d^{m-1}k\ \tilde S(k) \tilde S(-k)\exp \left( -t
\frac{k^2}{8}\right)I_{0}(t \frac{k^2}{8})
\,.\label{hk22}\end{equation}

This completes the calculation of the trace of the heat kernel to
the order of two reflections. We stress that equation (\ref{hk22}) is
exact at this order, as it contains also all non-asymptotic
contributions.

By going back to coordinate integrals, and expanding in $t$, we can
write a less compact but more explicit form of the heat kernel
asymptotics:
\begin{equation}
K_2=\sum_{l=0}^\infty \int_{\partial {\cal M}} dz\ S(z) \partial^{2l}
S(z) \frac{t^{-\frac{m-1}2 +1+2l}}{(4\pi )^{\frac {m-1}2}}
\frac{\Gamma (l+\frac 12)\,\Gamma (l+\frac 32)}{\pi l!\Gamma (2l+2)}
\,. \label{h2as}
\end{equation}
Note that, in contrast to (\ref{hk22}), this expansion is asymptotic,
and it cannot be used for large values of $t$.

\section{Effective boundary actions}

The results of the previous section can be used to calculate some
parts of the one-loop effective action induced on the boundary by the
bulk fluctuations. In what follows we will, in fact, obtain the full
one-loop effective action for a constant $S$ field, and its kinetic
piece in the case of an $S$ field depending on the boundary
coordinate.

In the $\zeta$-function regularization scheme, the renormalised
effective action is given by
\begin{equation}
W^{\mbox{\scriptsize reg}}=-\frac 12 \zeta^D(0)'\,. \label{renea}
\end{equation}
Here, $D$ is the operator defined by equations (\ref{quaction}) and
(\ref{rbc}).  The $\zeta$-function $\zeta^D(s)$ is the
Mellin transform of the trace of the heat kernel of the operator $D$:
\begin{equation}
\zeta^D (s)=\mu^{2s} {\rm Tr}(D^{-s})
= \frac {\mu^{2s}}{\Gamma (s)} \int_0^\infty
dt~t^{s-1} {\rm Tr} (\exp (-tD)) \,. \label{defzD}
\end{equation}
The parameter $\mu$ with the dimension of a mass has been introduced
to make the $\zeta$-function dimensionless. The resulting
 $\log (\mu )$ describes the
renormalization ambiguity.

In order to avoid infrared divergences, we will study the zeta
function for a massive bulk field, $M$ being its mass. As we will
see, the limit $M\rightarrow 0$ is direct in some cases, while an
analytic continuation of the zeta function is required before taking
such limit in other cases. The
effect of this nonzero mass ($M$) can easily be seen (for instance,
from equation (\ref{solint})) to reduce to multiplication of each
term in the trace of the heat kernel by $e^{-M^2 t}$.

\subsection{Coleman--Weinberg potential on the boundary}
\label{31}

The Coleman--Weinberg potential is, by definition, the effective
action for constant background fields. In this case, the expression
(\ref{nodir2}) applies for the trace of the heat kernel.

Due to translational invariance, an overall divergence proportional to the
volume of the boundary will appear in the effective action. This is
not, however, an obstacle to the approach, since it is the
Lagrangian density that has a physical meaning for Coleman--Weinberg
type effective actions.

Before going to the actual calculation, let us explicitly rewrite
the trace of the heat kernel for a massive bulk field

\begin{equation}
K(t)=\frac {V}{2(4\pi t)^{\frac{m-1}2}} \left( e^{(S^2-M^2)\,t} {\rm
erf} (S\sqrt t ) +e^{(S^2-M^2)\,t} \right)\,. \label{hkerfm}
\end{equation}

For $S^2<M^2$, $K(t)$ decays exponentially when $t\rightarrow\infty$.
On the other hand, for $S^2>M^2$, due to the asymptotic large-t
behaviour of the error function, $K(t)$ diverges exponentially in
this limit when $S>0$, and its Mellin transform can't be naively
performed. This is due to the presence of eigenmodes of the form
$e^{-Sr}$, which satisfy (\ref{rbc}) and fall off as $r\to +\infty$.
This modes eventually correspond to negative eigenvalues of
$-\partial^2 +M^2$, and their contribution to the $\zeta$ function
must be explicitly calculated and added to the Mellin transform of
the trace of the heat kernel, once the divergent contribution is
subtracted from the last. As a consequence, as we will see later,
for $S^2>M^2$ with $S>0$ the effective Lagrangian receives an
imaginary contribution, and presents a sign ambiguity. So, let us
consider the three different situations in a separate way.

\bigskip

a) $S^2<M^2$

\bigskip

The zeta function can be easily calculated by Mellin-transforming the
expression (\ref{hkerfm}) for the trace of the heat kernel. Before
integrating over the proper time $t$, it is convenient to write the
error function  as an integral:
\begin{equation}
e^{(S^2-M^2)\,t} {\rm erf} (S\sqrt t )=2S \sqrt{\frac t\pi}
\int\limits_0^1 d\xi\, e^{-(M^2+(\xi^2-1)S^2)\,t} \,.\label{interf}
\end{equation}

Thus,
\begin{equation}
K(t)=\frac {V}{2(4\pi t)^{\frac{m-1}2}} \left( 2S \sqrt{\frac t\pi}
\int\limits_0^1 d\xi\, e^{-(M^2+(\xi^2-1)S^2)\,t}
+e^{(S^2-M^2)\,t}\right)\,. \label{hkerint}
\end{equation}
In this case,  a closed expression for the $\zeta$-function in terms
of the hypergeometric function can be given. However, such an
expression is not very useful in practical calculations. By using
equations (\ref{renea}) and (\ref{hkerint}) the effective Lagrangian
for the case of a massive bulk field can be evaluated in a very fast
and efficient way in this case ($S^2<M^2$). In fact, since the
integral in (\ref{hkerint}) converges uniformly, the derivative with
respect to $s$ can be performed, and it can be evaluated at $s=0$
before actually doing the integral. The results for
$m=1,\,2,\,3,\,4,\,5$ are given in Appendix \ref{A3}. This case is
not relevant to the study of the problem with $M=0$.

\bigskip

b) $S^2>M^2$, $S>0$

\bigskip

As anticipated in the beginning of this Subsection, our first task
will be, in this case, to identify the exponentially growing part of
the trace of the heat kernel (\ref{hkerfm}). This can be done by
writing ${\rm erf} (S\sqrt t )=1-{\rm erfc} (S\sqrt t )$, where
\[{\rm erfc} (S\sqrt t )=
\frac{2}{\sqrt \pi}\int_{S\sqrt t}^{\infty}d\xi \,e^{-\xi^2}
\]
is the complementary error function. From its behaviour for large
$t$, it is clear that the piece to be subtracted from the trace of the
heat kernel is given by
\begin{equation}
K_{div}(t)=\frac{Ve^{(S^2-M^2)t}}{(4\pi t)^{\frac{m-1}2}}\,.
\label{exp}
\end{equation}

In Appendix \ref{A2} we show that, as we have already commented, this
piece comes from those modes of the Laplacian eventually becoming
negative. So, once this piece is subtracted, the trace of the heat
kernel reads
\begin{equation}
K_{sub}(t)=\frac {-V}{2(4\pi t)^{\frac{m-1}2}} \left(
e^{(S^2-M^2)\,t} {\rm erfc} (S\sqrt t )\right)\,. \label{hkersus}
\end{equation}

The Mellin transform of this quantity will give a first
contribution, $\zeta^D_1(s)$, to the relevant $\zeta$ function
($\zeta^D(s)$) . A second one will come from the explicit
contribution of the subtracted modes. Also in Appendix \ref{A2}, this
contribution is shown to be given by
\[
\zeta^D_2(s)=\frac{V\mu^{2s}(S^2-M^2)^{-s+\frac{m-1}{2}}}{(4\pi)^{\frac{m-1}{2}}\Gamma
\left(\frac{m-1}{2}\right)}\left[(-1)^{-s}\frac{\Gamma
\left(\frac{m-1}{2}\right)\Gamma \left(1-s\right)}{\Gamma
\left(\frac{m+1-2s}{2}\right)}+\right.\]
\begin{equation}\left.\frac{\Gamma
\left(s+\frac{1-m}{2}\right)\Gamma \left(1-s\right)}{\Gamma
\left(\frac{3-m}{2}\right)}\right]\,. \label{z22}
\end{equation}

As regards $\zeta^D_1(s)$, it can be obtained by Mellin-transforming
(\ref{hkersus}), with the complementary error function written in
integral form, and following the same steps as in the $S^2<M^2$ case.
It is given by

\begin{eqnarray}
&&\zeta^D_1(s)=\frac {-VS\mu^{2s}}{2(4\pi )^{\frac{m-1}2}}
\frac{\Gamma\left( s-\frac
m2+1\right)}{\Gamma(s)}\left[\frac{\left(S^2-M^2\right)^{\frac
m2-\frac12-s}}{\pi S}\Gamma \left( s-\frac m2 +\frac 12 \right)\times\right.\nonumber\\
&&\left.\Gamma \left(\frac m2-s
\right)-\frac{2M^{m-2s}}{S^2\pi^{\frac 12}(m-2s)}F\left
(\frac12\,,1\,,1+\frac m2 -s\,,\frac{M^2}{S^2}\right)\right]\,.
\label{z11}
\end{eqnarray}

From the total $\zeta$ function, which is the sum of (\ref{z22}) and
(\ref{z11}), the effective action for $S^2>M^2$ can be obtained for
any dimension $m$, and $S>0$. In the odd dimensional cases, where the
$\Gamma$ in the denominator supplies a power of $s$, an analytic
result for the effective Lagrangian can be obtained. However, for
even dimensions, the derivative with respect to $s$ of the
hypergeometric function can only be performed numerically. At
variance with case a), the derivative can't be evaluated at $s=0$
before performing the integral, since the last extends to a
non-compact interval. For completeness, we give the analytic result
for the effective Lagrangian in the case $m=1$ in Appendix \ref{A3}.

Let us now discuss the limit $M=0$. The limit of equation
(\ref{z22}) is direct. As regards equation (\ref{z11}), it is valid
for $\Re(s)>\frac{m-1}{2}$. The zero mass limit can be taken by
restricting, furthermore, to $\Re(s)<\frac{m}{2}$ (this is due to
the fact that the trace of the heat kernel behaves at infinity as a
power, rather than a decaying exponential). In this strip of the
$s$-plane, one has
\[
\zeta^D(s)\rfloor_{M=0}=\frac{V\mu^{2s}}
{(4\pi)^{\frac{m-1}{2}}}\left\{\frac{(S^2)^{-s+\frac{m-1}{2}}}{\Gamma
\left(\frac{m-1}{2}\right)}\left[(-1)^{-s}\frac{\Gamma
\left(\frac{m-1}{2}\right)\Gamma \left(1-s\right)}{\Gamma
\left(\frac{m+1-2s}{2}\right)}+\right.\right.\]
\[\left.\frac{\Gamma
\left(s+\frac{1-m}{2}\right)\Gamma \left(1-s\right)}{\Gamma
\left(\frac{3-m}{2}\right)}\right]-
\]
\begin{equation}
\left.S^{m-2s-1}\frac{\Gamma \left( s-\frac m2 +\frac12
\right)\Gamma \left( s-\frac m2 +1 \right)\Gamma \left( \frac m2 -s
\right)}{2\pi\Gamma (s)}\right\}\,.
\end{equation}

From this $\zeta$ function, the effective Lagrangian for $M=0$ and
$S>0$ can be easily obtained for any dimension. We list the results
for $m=1$ and $m=2$.

For $m=1$ we have:
\begin{equation}
{\cal L}^{\mbox{\scriptsize eff}}= -\frac12
\log\left(\frac{S^2}{\mu^2}\right)\pm i\pi\,.
\end{equation}

For $m=2$

\begin{equation}
{\cal L}^{\mbox{\scriptsize eff}}=
-\frac{S}{2\pi}\left(\log{\left(\frac{4S^2}{\mu^2}\right)}-2\right)\pm
\frac{iS}{2} \,,\end{equation} where $\gamma$ is Euler's constant,
and $\psi(x)$ is the zero order polygamma function.

As anticipated, a sign ambiguity arises in the effective Lagrangian,
due to the existence of negative eigenvalues.

\bigskip

c) $S^2>M^2$, $S<0$

\bigskip

In this case, the trace of the heat kernel behaves as $e^{-M^2\,t}$
times a power when $t\rightarrow\infty$, and a region in the $s$
plane exists where it can be Mellin-transformed. It is convenient to
write it as
\begin{equation}
K(t)=\frac {V}{2(4\pi t)^{\frac{m-1}2}} \left( e^{(S^2-M^2)\,t} {\rm
erfc} (|S|\sqrt t )\right)\,. \label{h}
\end{equation}

The resulting $\zeta$ function can be retrieved from (\ref{z11}) by
changing the overall sign and turning $S$ into $|S|$. Thus, it is
given by

\begin{eqnarray}
&&\zeta^D(s)=\frac {V|S|\mu^{2s}}{2(4\pi )^{\frac{m-1}2}}
\frac{\Gamma\left( s-\frac
m2+1\right)}{\Gamma(s)}\left[\frac{\left(S^2-M^2\right)^{\frac
m2-\frac12-s}}{\pi |S|}\Gamma \left( s-\frac m2 +\frac 12 \right)\times\right.\nonumber\\
&&\left.\Gamma \left(\frac m2-s
\right)-\frac{2M^{m-2s}}{S^2\pi^{\frac 12}(m-2s)}F\left
(\frac12\,,1\,,1+\frac m2 -s\,,\frac{M^2}{S^2}\right)\right]\,.
\end{eqnarray}

The comments made after equation (\ref{z11}) also apply to this
case. The effective Lagrangian in the massive case is given, for
$m=1$ in Appendix \ref{A3}.

In the limit $M=0$ one obtains, for $S<0$
\begin{equation}
\zeta^D(s)\rfloor_{M=0}=\frac{V\mu^{2s}}
{2(4\pi)^{\frac{m-1}{2}}}|S|^{m-2s-1}\frac{\Gamma \left( s-\frac m2
+\frac12 \right)\Gamma \left( s-\frac m2 +1 \right)\Gamma \left(
\frac m2 -s \right)}{\pi\Gamma (s)}\,.
\end{equation}

and the effective actions for $m=1$ and $m=2$ are as follows

For $m=1$ we have:
\begin{equation}
{\cal L}^{\mbox{\scriptsize eff}}= -\frac12 \log \left( \frac
{S^2}{\mu^2} \right)\,.
\end{equation}

For $m=2$

\begin{equation}
{\cal L}^{\mbox{\scriptsize eff}}=
\frac{|S|}{2\pi}\left(\log{\left(\frac{4S^2}{\mu^2}\right)}-2\right)\,.
\end{equation}

Let us discuss some qualitative features of the potentials obtained
above. These potentials have a very non-trivial structure, especially
for $M\ne 0$, providing a lot of possibilities for the symmetry
breaking with the formation of a boundary condensate. The presence of
an imaginary part for $S>M$ indicates an instability of the ``Robin
phase'', which may eventually decay to some other (Dirichlet?) phase.
Here we see many similarities with the tachyon condensation and
$D$-brane formation in open string theory. We are going to address
these questions in a separate publication.

\subsection{Effective kinetic energy}
\label{32}

In this section, we will evaluate the one-loop correction to the
kinetic energy of the boundary field $S(z)$, due to the quantum
fluctuations of the bulk scalar field $\phi$. Since we are
interested only in the kinetic part of the effective action (or,
equivalently, the propagator), it will be enough to consider the
multiple reflection expansion to the order of two reflections
(higher orders will contain higher powers of the field $S$).

The term involving no reflection at the boundary will be ignored. In
fact, it is $S$-independent and can always be eliminated through a
redefinition of the cosmological constant on the brane.

The one-reflection contribution to the zeta function is seen, from
equation (\ref{hk1}), to be given by
\[
\zeta_1= \frac{\mu^{m-2}2^{1-m}}{\pi^\frac{m}{2}\Gamma(s)}\tilde
S(\tilde k=0)\int_0^{\infty}dt t^{s+\frac {(2-m)}{2}-1}
e^{-\frac{M^2}{\mu^2} t}\]
\begin{equation}=\frac{\mu^{m-2}\Gamma(s-\frac
{(m-2)}{2})2^{1-m}}{\pi^\frac{m}{2}\Gamma(s)}\tilde S(\tilde
k=0){(\frac{M^2}{\mu^2})}^{(\frac {(m-2)}{2}-s)},\, for\,
\Re(s)>\frac{m-2}{2}\,, \label{z1}\end{equation} which can be
meromorphically extended to the region $\Re(s)<\frac{m-2}{2}$. The
limit $M\rightarrow 0$ can be seen to vanish in this region. So, no
contribution from one reflection will appear in the effective
action, for any dimension $m$.

Let us now go to the 2-reflection contribution. From
equation(\ref{hk22}) the zeta function is seen to be given by

\[ \frac{\mu^{m-3}}{2(4\pi)^{\frac{m-1}2}\Gamma(s)}
 \int d^{m-1}k\ \tilde S(k)
\tilde S(-k)\int_0^{\infty}dt{t^{s+\frac{3-m}{2}-1}}\exp \left( -t
(\frac{k^2}{8\mu^2}+\frac{M^2}{\mu^2}) \right)I_{0}(t
\frac{k^2}{8\mu^2})\]
\[
=\frac{\mu^{m-3}\Gamma(s+\frac{3-m}{2})}{2(4\pi)^{\frac{m-1}2}\Gamma(s)}
 \int d^{m-1}k\ \tilde S(k)
\tilde S(-k)\left(\frac{k^2}{8\mu^2}\right)^{-s+\frac{m-3}{2}}
\left(1+\frac{8M^2}{k^2}\right)^{\frac{m-3}{2}-s}\times\]
\begin{equation}F(\frac{s}{2}+\frac{5-m}{4},\frac{s}{2}
+\frac{3-m}{4};1;\frac{k^4}{(k^2+8M^2)^2}), \,for\,
\Re(s)>\frac{m-3}{2}\,, \label{z2}\end{equation} where $F$ is the
hypergeometric function. Now, from the properties of these
functions, it is easy to see that it is only for $m=2$ that the
analytic extension must be performed before the zero mass limit. For
$m\geq 3$, $M$ can be taken to zero from the beginning.

Let us look at two interesting cases:

a) $m=2$

In this case

\[
\zeta_2=\frac{\mu^{-1}\Gamma(s+\frac{1}{2})}{2(4\pi)^{\frac{1}2}\Gamma(s)}
 \int d^{m-1}k\ \tilde S(k)
\tilde S(-k)\left(\frac{k^2}{8\mu^2 }\right)^{-s-\frac{1}{2}}
\left(1+\frac{8M^2}{k^2}\right)^{-\frac{1}{2}-s}\times\]
\[\left[\frac{\Gamma(-s)}{\Gamma(\frac14-\frac{s}{2})
\Gamma(\frac34-\frac{s}{2})}F(\frac{s}{2}+\frac{3}{4},\frac{s}{2}
+\frac{1}{4};s;1-\frac{k^4}{(k^2+8M^2)^2})+\right.\]
\begin{equation}
\left.\left(\frac{-8M^2}{k^2}\right)^{-s}
\frac{\Gamma(s)}{\Gamma(\frac14+\frac{s}{2})
\Gamma(\frac34+\frac{s}{2})}F(-\frac{s}{2}+\frac{3}{4},-\frac{s}{2}
+\frac{1}{4};1-s;1-\frac{k^4}{(k^2+8M^2)^2})\right]\,.
\end{equation}

Now, for $\frac{-1}{2}<\Re(s)<0$, the second term inside the square
brackets vanishes when $M=0$, and one gets
\begin{equation}
\zeta_2=\frac{2^{2s-2}\mu^{-1}\Gamma(s+\frac{1}{2})\Gamma(-s)}{\pi\Gamma(s)\Gamma(\frac12-s)}
 \int dk\ \tilde S(k)
\tilde S(-k)\left(\frac{k^2}{\mu^2}\right)^{-s-\frac{1}{2}}\,.
\end{equation}

The effective action is then seen to be given by
\begin{equation}
W^{\mbox{\scriptsize eff}}=\frac{1}{2\pi}\int dk\ \tilde S(k) \tilde
S(-k)\left(k^2\right)^{-\frac{1}{2}}\left[2\gamma+2\log{2}+2\psi\left(\frac12
\right) -\log(\frac{k^2}{\mu^2})\right]\,. \label{S2m2}
\end{equation}

Since $\zeta_2(s=0)\neq 0$, a dependence on $\log(\mu)$ remains.
Comparing with (\ref{h2as}) we see that there is no term of this
form in the small $t$ asymptotics of the heat kernel. We conclude
that the $\log(\mu)$ term appears due the infrared behavior of the
heat kernel (large $t$). $\mu$ thus plays the role of an infrared
regulator, rather than a renormalization parameter. The
corresponding ambiguity cannot be fixed by a normalization condition
because this would require the presence of a nonlocal
($\partial_z^{-1}$) counter term. Usually, such problems are solved
by a resummation of the perturbation theory series. The IR
singularity $\mu\to 0$ in (\ref{S2m2}) looks similar to the one
which appears in two dimensions on manifolds without boundary. It
has been demonstrated in \cite{Gusev:2000cv} that for this latter
case all infrared problems disappear if one uses the dilaton
representation \cite{Kummer:1999dc} for the potential and then
collect powers of the dilaton instead of powers of the potential
itself.

b) $m=5$

Here, the limit $M=0$ can be taken directly in equation (\ref{z2}),
and one easily gets
\[
\zeta_5=\frac{\mu^{2}}{2(4\pi)^{2}(s-1)}
 \int d^{4}k\ \tilde S(k)
\tilde S(-k)\left(\frac{k^2}{8\mu^2}\right)^{-s+1}
F(\frac{s}{2},\frac{s-1}{2};1;1)=\]
\begin{equation}
\frac{\mu^{2}}{2^7(\pi)^{\frac52}(s-1)}\frac{\Gamma(\frac32-s)}{\Gamma(2-s)}
 \int d^{4}k\ \tilde S(k)
\tilde S(-k)\left(\frac{k^2}{\mu^2}\right)^{-s+1} \,.
\end{equation}
The effective action can be easily obtained, and it is given by
\begin{equation}
W^{\mbox{\scriptsize eff}}=
-\frac{1}{2^8(\pi)^{\frac52}}\int d^{4}k\ \tilde S(k) \tilde
S(-k)\,k^2\left(2-\gamma-\psi\left(\frac32
\right)-\log\left(\frac{k^2}{\mu^2}\right)\right)\,.\label{47}
\end{equation}
This effective action also depends on $\log(\mu)$. But his time,
$\log(\mu)$ appears in front of a local action $S\Delta S$. Such
term is present in the small $t$ asymptotics of the heat kernel.
Therefore, the $\mu$-dependence appears due to ultra violet
divergences. In principle, the dependence on $\mu$ can be
renormalised away provided there is a suitable term in the classical
action.

As we have noted after eq. (\ref{mcon}), our results for $m=5$ can be
used also in a quantum brane world scenario. Suppose that the scalar
field potential on the brane has the usual Higgs form: $\tilde
V(\phi )\sim \phi^4$. $S$ has to be identified with the second
derivative of this potential, $S\sim \bar\phi^2$. Hence, $S\Delta
S\sim \bar\phi^2\Delta\bar\phi^2 $. Renormalising this term would
require a rather unusual interaction term in the classical action.
This is just another example of exotic counterterms
\cite{Gilkey:2001mj} which appear in the brane-world scenario. The
finite part of the quantum correction goes as $k^2 \log(k^2)$.
Already on dimensional grounds, it is clear that typical scalar
theories in four dimensions do not present such strong growth of the
effective action at large momenta. Therefore, we come to the (not
unexpected) conclusion that the short distance quantum physics
depends strongly on the presence of extra dimensions. For more
realistic brane models in curved space the heat kernel expansion
\cite{Gilkey:2001mj} predicts other $S^2$-terms proportional to
geometric characteristics of the brane (but independent of $k$).
These terms will dominate (\ref{47}) at small momenta.

Note that in any even dimension greater than 2, $\zeta(0)=0$. Thus,
no logarithmic operator or dependence on $\mu$ appear in the
effective action.

\section{Conclusions}
In this paper we used the multiple reflection expansion to calculate
(parts of) the heat kernel and of the effective action depending
on the function $S$ appearing in Robin boundary conditions (\ref{rbc}).
In particular, we have calculated the Coleman-Weinberg potential
on the boundary (assuming constant $S$) and the quadratic part
in $S$ containing arbitrary number of tangential derivatives.
Applications to tachyon condensation in open string theory and to
the brane-world scenario were briefly outlined. More detailed study
of these applications will be given elsewhere.

The basic relations (\ref{Deq}) and (\ref{solint}) of the multiple
reflection expansion method can be extended to other boundary value
problems, which correspond to taking more complicated operators
instead of just the scalar function $S$. Acting in this way we may
cover boundary conditions appearing in the context of Casimir energy
calculations \cite{Bordag:2001qi}, solid state physics and strings
(see \cite{Bordag:2000ux,Vassilevich:2002wf}  for some examples).

\section*{Acknowledgements}

The authors acknowledge support from Fundación Antorchas and DAAD
(grant 13887/1-87). The work of DVV has been supported by the DFG
project Bo 1112/11-1. HF and EMS also acknowledge CONICET (Argentina)
(grant 0459/98).
\appendix

\section{Curved boundary}
\label{A1}

In this appendix we demonstrate that our results can be extended to
the case of a curved boundary. To this end, we use the conformal
variation technique, which is quite different from the expansions
used in the main text. For simplicity, we neglect derivatives of $S$.

The heat kernel coefficients for Robin boundary conditions are
locally computable. This means that they can be expressed through
volume and surface integrals of some local invariants. In
particular, on dimensional grounds, we can write
\begin{equation}
a_n(f,D)\simeq (4\pi )^{-(m-1)/2}\int_{\partial {\cal M}}dz\, \left(
c_0 S^{n-1}+c_1f_{;m}S^{n-2}+ c_2L_{aa}S^{n-2}+\dots \right)
\,,\label{genan}
\end{equation}
where $L_{aa}$ denotes the trace of the extrinsic curvature of the
boundary. We dropped many other invariants which are not relevant
for the present calculation. It is important that the constants
$c_k$ do not depend on $m$.

Let us consider a local conformal transformation of the operator $D$:
$D\to e^{-2\epsilon f}D$. All local invariants and the heat kernel
coefficients change under this transformation. One can show
\cite{BG90} that
\begin{equation}
\frac d{d\epsilon} \vert_{\epsilon =0} a_n (1,e^{-2\epsilon
f}D)=(m-n)a_n(f,D)\,. \label{confrel}
\end{equation}
Transformation rules for individual local invariants can be found in
\cite{BG90} (see also \cite{Vassilevich:1995we}). Let us consider
the terms which produce $f_{;m}S^{n-2}$ ($n\ge 3$) after conformal
variation. Obviously, there are only two such terms:
\begin{eqnarray}
&&\frac d{d\epsilon} \vert_{\epsilon =0} S^{n-1} \to
\frac{(m-2)(n-1)}2 f_{;m}S^{n-2} \nonumber \\
&&\frac d{d\epsilon} \vert_{\epsilon =0} L_{aa}S^{n-2} \to
-(m-1)f_{;m}S^{n-2} \label{Sfm}
\end{eqnarray}
One has to remember that the volume element on the boundary is also
changed: $dz\to e^{(m-1)\epsilon f}dz$.

Collecting the terms containing $f_{;m}S^{n-2}$ on both sides of
equation (\ref{confrel}), we obtain ($n>2$):
\begin{equation}
\frac{n-1}2 (m-2)c_0-(m-1)c_2=(m-n)c_1 \,.\label{ceq}
\end{equation}
The equation (\ref{ceq}) can be solved giving:
\begin{equation}
c_1=c_0/2 \,.\label{c1}
\end{equation}

We can immediately calculate all terms in the heat kernel expansion
which are linear in $L_{ab}$ and contain arbitrary power of $S$:
\begin{equation}
K(f,t)\simeq\frac 1{(4\pi )^{\frac{m-1}2}} \int_{\partial {\cal M}}
dz f(z) \sum\limits_l S^l(z)L_{aa}(z) t^{\frac 12 (l-m+2)}
\frac{l}{4\Gamma \left( \frac{l+3}2  \right)} \label{laa}
\end{equation}
This is consistent with \cite{BG90,Kirsten:1998qd} for $l=1,2,3$. As
an additional consistency check we see that all dependence on $m$
resides in a power of $4\pi t$ only.

The right hand side of (\ref{laa}) can again be represented through
the error function ( c.f. eq. (\ref{hkerf})). All the calculations of
Section \ref{31} can be repeated step by step, thus giving the
effective potential on a slightly curved boundary.

\section{Discussion of negative eigenvalues}
\label{A2}

We will, in the first place, prove that the exponentially growing
part (\ref{exp}) of the trace of the heat kernel for $S^2>M^2$ is
precisely the contribution due to the eigenvalues of
$-{\partial}^2+M^2$ eventually becoming negative. Such eigenvalues
are of the form
\[
\lambda_{-}=M^2-S^2+k^2\,,
\]
where $k$ is the boundary momentum. Then, their contribution to the
trace of the heat kernel is given by
\[
K_{div}(t)=\frac{V}{(2\pi)^{m-1}}\int_{-\infty}^{\infty}d^{m-1}k\,e^{-(M^2-S^2+k^2)t}\,,
\]
where the prefactor is the density of states on the boundary.

After making explicit the integration measure, one has
\[
K_{div}(t)=\frac{2V\pi^{\frac{m-1}{2}}}{(2\pi)^{m-1}\Gamma
\left(\frac{m-1}{2}\right)}\int_{0}^{\infty}dk\,k^{m-2}\,e^{-(M^2-S^2+k^2)t}\,.
\]

By changing the integration variable to $k^{'}=k^2t$, the final
result arises, which is
\begin{equation}
K_{div}(t)=\frac{Ve^{(S^2-M^2)\,t}}{(4\pi t)^{\frac{m-1}{2}}} \,.
\label{kdiv}
\end{equation}

Now, we will determine the contribution of these modes to the zeta
function in a direct way. It is given by
\[\zeta^D_2(s)=\frac{2V\mu^{2s}\pi^{\frac{m-1}{2}}}{(2\pi)^{m-1}\Gamma
\left(\frac{m-1}{2}\right)}\int_{0}^{\infty}dk\,k^{m-2}\left(M^2-S^2+k^2\right)^{-s}\,,
\]
or, changing variables,
\[\zeta^D_2(s)=\frac{V\mu^{2s}}{(4\pi)^{\frac{m-1}{2}}\Gamma
\left(\frac{m-1}{2}\right)}\left[(-1)^{-s}\int_{0}^{S^2-M^2}dk\,k^{\frac{m-3}{2}}
\left(S^2-M^2-k\right)^{-s}+\right.\]\[\left.\int_{S^2-M^2}^{\infty}dk\,k^{\frac{m-3}{2}}
\left(M^2-S^2+k\right)^{-s}\right]\,.
\]

After performing the integrals, one finally gets
\[
\zeta^D_2(s)=\frac{V\mu^{2s}(S^2-M^2)^{-s+\frac{m-1}{2}}}{(4\pi)^{\frac{m-1}{2}}\Gamma
\left(\frac{m-1}{2}\right)}\left[(-1)^{-s}\frac{\Gamma
\left(\frac{m-1}{2}\right)\Gamma \left(1-s\right)}{\Gamma
\left(\frac{m+1-2s}{2}\right)}+\right.\]
\begin{equation}\left.\frac{\Gamma
\left(s+\frac{1-m}{2}\right)\Gamma \left(1-s\right)}{\Gamma
\left(\frac{3-m}{2}\right)}\right]\,.
\end{equation}

\section{Coleman-Weinberg Lagrangians for $M\neq 0$}
\label{A3}

a) $S^2<M^2$

For $m=1$
\begin{equation}
{\cal L}^{\mbox{\scriptsize
eff}}=\frac{1}{2}\log\left({\frac{M-S}{\mu}}\right)\,.
\end{equation}
The dependence on $\mu$ can be renormalised away by requiring that
the effective Lagrangian vanishes when $M\rightarrow \infty$. In
this case, this is equivalent to subtracting a field-independent
term, which is merely a redefinition of the cosmological constant.
After doing so, one gets

\begin{equation}
{\cal L}^{\mbox{\scriptsize eff}}=\frac 12 \log \left( 1-\frac SM
\right) \,,
\end{equation}

For $m=2$

\[
{\cal L}^{\mbox{\scriptsize eff}}=\frac{S}{4\pi}
\left(-2+\log{\frac{M^2}{\mu^2}}+2\frac{\sqrt{M^2-S^2}}{S}\arctan{\left(\frac{S}
{\sqrt{M^2-S^2}}\right)}\right)\]
\begin{equation}+\frac14 \sqrt{M^2-S^2}\,.
\end{equation}

The dependence on $\mu$ can again be eliminated by asking the
effective Lagrangian to vanish in the infinite mass limit. But, in
this case, this amounts to renormalizing a term linear in $S$.

For $m=3$

\begin{equation}
{\cal L}^{\mbox{\scriptsize eff}}=\frac{1}{16\pi}
\left(M^2-S^2+2MS-2\left(M^2-S^2\right)\log\left(\frac{M-S}{\mu}\right)\right)
\,.
\end{equation}

For $m=4$

\[
{\cal L}^{\mbox{\scriptsize eff}}=\frac{-1}{144\pi^2} \left(16
S^3-21 M^2 S+6\pi\left(M^2-S^2\right)^{\frac32}
+6\left(3M^2S-2S^3\right)\log\left({\frac{M}{\mu}}\right)+\right.\]
\begin{equation}
\left.12\left(M^2-S^2\right)^{\frac32} \arctan{\left(\frac{S}
{\sqrt{M^2-S^2}}\right)}\right) \,.
\end{equation}

For $m=5$
\begin{eqnarray}
&&{\cal L}^{\mbox{\scriptsize eff}}=
\frac{1}{768\, {\pi }^2} \left( -9\, M^4 - 20\, M^3\, S
+ 18\, M^2\, S^2 + 12\, M\,
   S^3 - 9\, S^4 \right.
\\ \nonumber && \qquad \left.
+ 12\, {\left( M^2 - S^2 \right) }^2\, \log (M - S) \right)
\end{eqnarray}

\bigskip

b) $S^2>M^2$, $S>0$

For $m=1$

\begin{equation}
{\cal L}^{\mbox{\scriptsize eff}}=-\frac12
\log\left(\frac{S^2-M^2}{\mu^2}\right)+\arg\tanh{\left(\frac{M}{S}\right)}\pm
i\pi\ \,.
\end{equation}

c) $S^2>M^2$, $S<0$

For $m=1$

\begin{equation}
{\cal L}^{\mbox{\scriptsize eff}}=-\frac12
\log\left(\frac{S^2-M^2}{\mu^2}\right)-\arg\tanh{\left(\frac{M}{|S|}\right)}\,.
\end{equation}

\end{document}